\documentclass[10pt,prl,twocolumn,amsfonts,amssymb,amsmath,aps,notitlepage]{revtex4-2}
\usepackage{graphicx}
\usepackage{subfigure}
\usepackage{amsmath}
\usepackage{color}
\usepackage{hyperref}

\def\ket#1{\lvert\nobreak#1\nobreak\rangle}
\def\bra#1{\langle\nobreak#1\nobreak\rvert}

\def\adag{\hat{a}^{\dag}{}}
\def\a{\hat{a}}

\def\alphamag{\lvert \alpha \rvert}
\def\sigmap{\hat{\sigma}_+}
\def\sigmam{\hat{\sigma}_-}
\def\sigmaz{\hat{\sigma}_z}

\def\Psipket{\ket{\Psi_+(t)}}
\def\Psimket{\ket{\Psi_-(t)}}
\def\Psipbra{\bra{\Psi_+(t)}}
\def\Psimbra{\bra{\Psi_-(t)}}

\newcommand{\op}[1]{\hat{#1}}
\newcommand{\opdag}[1]{\hat{#1}^{\dag}{}}
\newcommand{\abs}[1]{\lvert #1 \rvert}

\begin{document}

\title{Quantum-corrected Floquet dynamics: bridging fully quantum and semiclassical regimes}
\author{E. K. Twyeffort}
\affiliation{School of Physics and Astronomy, University of Southampton, Highfield, Southampton, SO17 1BJ, United Kingdom}
\email{e.k.twyeffort@soton.ac.uk}
\author{A. D. Armour}
\affiliation{School of Physics and Astronomy and Centre for the Mathematics and Theoretical Physics of Quantum Non-Equilibrium Systems, University of Nottingham, Nottingham NG7 2RD, United Kingdom}

\date{\today}

\begin{abstract}
Semiclassical descriptions of a few-level system coupled to an electromagnetic field mode reduce the field to a time-dependent driving term. Although such methods are widely used, the underlying quantum character of the field generates corrections to the dynamics that can become significant. Here we develop a general approach for systematically calculating quantum corrections to the time-dependent Floquet dynamics that emerges in the semiclassical limit. Taking the Rabi model of a spin--field system as an example, we show how approximate analytical expressions for short-time quantum corrections to the semiclassical dynamics can be obtained. This framework helps explain the emergence of semiclassical behavior, sheds new light on collapse and revival dynamics in the Rabi model, and provides a tool for assessing corrections to semiclassical control techniques. 
\end{abstract}

\maketitle

The interaction of electromagnetic fields with few-level quantum systems --- spins, atoms, superconducting qubits, quantum dots --- underpins many emerging quantum technologies\,\cite{Saffman2010,Vandersypen2019,Chatterjee2021}. Increasingly sophisticated control techniques have been developed under the assumption that the fields may be treated classically\,\cite{Viola1998,Brif2010,Ivakhnenko2022}. Nevertheless, such fields are quantized and the dynamics they give rise to will not generally be the same as that of the corresponding semiclassical system. Indeed, even when the field is prepared in the most classical of quantum states, corrections to the semiclassical dynamics can lead over time to dramatically different behavior~\cite{Knight1981}. Understanding the nature and magnitude of quantum corrections to semiclassical dynamics is a problem of fundamental interest with practical significance for quantum control algorithms\,\cite{Armata2016}. Unlike other forms of noise and decoherence that can be mitigated, 
quantization of the field is an irreducible source of error in classical control protocols.

Intensive efforts over several decades have sought effective means of calculating the fully quantum dynamics of few-level systems coupled to quantum fields. Many studies have relied on numerical computation, as analytical approximation techniques for calculating dynamical variables are complicated and provide limited physical insight\,\cite{Narozhny1981, Miller1992, Fleischhauer1993, Feranchuk2009, Yan2024}. Although this work has uncovered the rich phenomenology present in even the simplest models, it has not managed to construct a complete, coherent picture of how quantum field dynamics can give rise to semiclassical behavior. The series of studies on quasiclassical trajectories 
by Gea-Banacloche~\cite{GeaBanacloche1991, GeaBanacloche1992, GeaBanacloche1992b}, together with Finney~\cite{Finney1994}, comes closest. Their work bears similarities to that presented here, but is rooted in phenomenological observations rather than the more general first-principles approach that we take.


In this Letter we introduce `quantum-corrected Floquet dynamics' (QCFD): a formalism that expresses the time evolution of the joint state of a quantum system interacting with a quantized field in terms of the Floquet solutions to the corresponding semiclassical Hamiltonian. We motivate and outline an approximation strategy for lowest-order quantum corrections to the semiclassical dynamics, the Floquet-basis rotating-wave approximation (FBRWA). Applying this approach to the paradigmatic Rabi model, we show that the FBRWA yields a closed-form analytical expression for the evolution of the full quantum state vector with an arbitrary initial state of the quantum field. To illustrate the power of this technique we derive excited-state population dynamics in both the Rabi model and the simpler Jaynes-Cummings model (JCM) and find new analytical results for the collapse dynamics of displaced Fock states and their superpositions. The FBRWA expressions accurately describe the collapse of the Rabi oscillations but not their later revival, which has an intuitive physical explanation.

The basis of QCFD is a division of the dynamics induced by the field into semiclassical and quantum components. This enables us to cleanly delineate which features in the evolution are semiclassical in nature and which arise from the underlying quantum character of the field. Consider a quantum system $S$ with a discrete spectrum, described by the Hamiltonian $H_S$, and interacting with a single-mode quantum field of frequency $\omega_0$:
\begin{equation}
    H = H_S + \omega_0 \adag \a + \sum_{m,n} \lambda_{mn} \ket{S_m} \bra{S_n} \otimes f_{mn}(\a, \adag) ,
\label{H_orig}
\end{equation}
where $\adag$ and $\a$ are the raising and lowering operators for the field and the state vectors $\{\ket{S_m}\}$ span the Hilbert space of $S$. The energy eigenstates of the field are the Fock states $\ket{n}$. The quantum coupling strengths $\lambda_{mn}$ and the function $f_{mn}(\a, \adag)$ define the interaction of $S$ with the field.

To separate the Hamiltonian into semiclassical and quantum terms, we draw on insight from the mathematical formalism for the quantum-to-semiclassical transition established in \cite{TwyIrish2022}. Moving to a rotating frame with respect to the field, followed by a displacement transformation $\op{D}(\alpha)$ on the field operators\,\cite{TwyIrish2022}, transforms $H$ into
\begin{equation}
\begin{split}
    \tilde{H}^{I}(t) = H_S + \sum_{m,n} &\lambda_{mn} \ket{S_m} \bra{S_n} \\
    &\otimes f_{mn}[(\a+\alpha)e^{-i \omega_0 t}, (\adag+\alpha^*)e^{i \omega_0 t} ].
\label{H_transf}
\end{split}
\end{equation}
For interactions that can be expressed as a power series, $f_{mn}(\a, \adag) = \sum_{j,k=0}^\infty c_{mn}^{jk} (\adag)^j \a^k$,
the transformed Hamiltonian may be written in the form 
\begin{equation}
        \tilde{H}^{I}(t) = H_{sc}(t)\otimes \op{I}_f + H_q^I(t) ,
\label{displ_transformb}
\end{equation}
where 
\begin{equation}
    H_{sc}(t) \equiv H_S + \sum_{mn} \lambda_{mn} \ket{S_m} \bra{S_n} \otimes f_{mn}(\alpha e^{-i \omega_0 t}, \alpha^* e^{i \omega_0 t} )
    \label{Hsc}
\end{equation} 
represents an effective time-dependent Hamiltonian for $S$, independent of the quantum state of the field mode. $\op{I}_f$ is the identity for the field and $H_q^I(t)$ describes the interaction with the quantized field~\footnote{Note that $H_{sc}(t)$ may differ from the semiclassical Hamiltonian obtained by applying the joint scaling limit $\lambda_{mn} \to 0$ and $\lvert \alpha \rvert \to \infty$ to Eq.~\eqref{H_transf}. An example is found in \cite{TwyIrish2022}, in which the quantum interaction generates a renormalization of the spin frequency that vanishes in the limit. This effect can be viewed as a zeroth-order quantum correction to the semiclassical dynamics and is easily accommodated in the QCFD formalism.}.

The effective Hamiltonian $H_{sc}(t)$ is periodic in time and hence admits Floquet solutions
\begin{equation}
\ket{\Psi_{m}(t)} = e^{-iq_{m}t} \ket{\psi_{m}(t)}
\label{floquet_sols}
\end{equation}
that satisfy
\begin{equation}
   \left[H_{sc}(t) - i \tfrac{\partial}{\partial t}\right]\ket{\psi_{m}(t)} = q_{m} \ket{\psi_{m}(t)} ,
\label{floquet_Sch_eq}
\end{equation}
where the Floquet quasienergies $q_{m}$ are defined up to an integer multiple of $\omega_0$\,\cite{Shirley1963,Finney1994}. These Floquet states form a complete orthonormal basis at any time $t$. An arbitrary state of the joint quantum system may be expanded as 
\begin{equation}
    \ket{\Phi(t)} = \sum_m c_m \ket{\Psi_m(t)} \otimes \ket{\tilde{\phi}_m^I(t)} .
\end{equation}
The field states are denoted in the transformed basis as $\ket{\tilde{\phi}_{m}^I(t)} = \sum_{n=0}^{\infty}d_{m}^n(t)\opdag{D}(\alpha) e^{i \omega_0 t \adag \a} \ket{n}$, with probability amplitudes $d_{m}^n(t)$. Inserting this form of the wavefunction into the Schr{\"o}dinger equation and projecting onto the Floquet states yields a set of coupled differential equations for the quantum field associated with each Floquet state of $S$:
\begin{equation}
    i \frac{\partial}{\partial t} \ket{\tilde{\phi}_m^I(t)} = \sum_{n} \bra{\Psi_{m}(t)} H_q^I(t) \ket{\Psi_{n}(t)} \ket{\tilde{\phi}_n^I(t)} .
\label{field_diffeqs}
\end{equation}
This is the first key conceptual result of QCFD. Working in the Floquet basis effectively diagonalizes the semiclassical term in the Hamiltonian, allowing corrections due to the quantum nature of the field to be isolated. 

This approach introduces explicit time dependence in both the Floquet states $\ket{\Psi_m(t)}$ and the quantum Hamiltonian $H_q^I(t)$. Using the definition \eqref{floquet_sols}, Eq.~\eqref{field_diffeqs} can be written as
\begin{equation}
\begin{split}
    i \frac{\partial}{\partial t} \ket{\tilde{\phi}_m^I(t)} &= \bra{\psi_{m}(t)} H_q^I(t) \ket{\psi_{m}(t)} \ket{\tilde{\phi}_m^I(t)} \\
    &+ \sum_{n \neq m} e^{-i(q_n-q_m)t} \bra{\psi_{m}(t)} H_q^I(t) \ket{\psi_{n}(t)} \ket{\tilde{\phi}_n^I(t)} .
\label{field_diffeqs2}
\end{split}
\end{equation}
Both $\ket{\psi_m(t)}$ and $H_q^I(t)$ are periodic in $\omega_0$ and can be expressed as Fourier series. Consequently, the diagonal elements of $H_q^I(t)$ in~\eqref{field_diffeqs2}  only involve powers of $e^{i \omega_0 t}$, while the off-diagonal terms contain the additional factor $e^{-i(q_n - q_m)t}$. The two terms on the left-hand side of Eq.~\eqref{field_diffeqs2} are distinct in physical character. The first leaves the Floquet states of $S$ unchanged but generates quantum evolution of the associated field components; the second creates transitions between Floquet states. 

The separation of timescales between the diagonal and off-diagonal terms of Eq.~\eqref{field_diffeqs2} invites an approximation strategy in the spirit of the standard rotating-wave approximation. Neglecting fast-rotating off-diagonal terms reduces \eqref{field_diffeqs2} to a set of uncoupled differential equations for the field. The further approximation of retaining only the time-independent component in the Fourier series representation of $\bra{\psi_{m}(t)} H_q^I(t) \ket{\psi_{m}(t)}$ (assuming it exists) immediately enables \eqref{field_diffeqs2} to be integrated, yielding a propagator for the time evolution of an arbitrary initial state of the field. We term this the Floquet-basis rotating-wave approximation (FBRWA).

As a concrete example, take the Rabi model: a two-level quantum system linearly coupled to a single mode of the electromagnetic field\,\cite{Braak2017,Xie2017,TwyIrish2022}. For simplicity, we use the term `spin' for the two-level system. We write the Hamiltonian in the form
\begin{equation}
    H = \tfrac{1}{2} \Omega \sigmaz + \omega_0 \adag \a + \lambda[f(\a, \adag) \sigmap + f^{\dag}(\a, \adag) \sigmam],
\label{H_orig}
\end{equation}
where $\op{\sigma}_{\pm} = \ket{\pm z}\bra{\mp z}$ are the raising and lowering operators for the spin eigenstates $\ket{\pm z}$. With the choice $f(\a, \adag) = (\a + \adag)$, Eq.~\eqref{H_orig} corresponds to the quantum Rabi model. Under the standard rotating-wave approximation (RWA), $f(\a, \adag) = \a$ and we recover the Jaynes-Cummings model (JCM).

With the choice $q_- = -q_+$, the Floquet states have the Fourier series representation \cite{Finney1994,Autler1955}\footnote{Note that the $B$ coefficients defined here are the negative of those given in \cite{Finney1994}.} 
\begin{equation}
    \ket{\Psi_{\pm}(t)} = e^{-i q_{\pm} t} (\pm A_{\pm}(t) \ket{\pm z} + B_{\pm}(t) \ket{\mp z} ),
\end{equation}
where $A_{\pm}(t) = \sum_{k=-\infty}^{\infty} A_{2k} e^{\pm 2ki\omega_0 t}$ and $B{_\pm} = \sum_{l=-\infty}^{\infty} B_{2l+1} e^{\pm (2l+1)i\omega_0 t}$ with $A_{2k}$, $B_{2l+1}$ real.

Expanding $H_q^I(t)$ in this basis (see Appendix for detailed expressions) and keeping only the time-independent terms yields the FBRWA interaction Hamiltonian, which takes the same remarkably simple form for both the JCM and the full Rabi model:
\begin{equation}
    H_{\rm{FBRWA}}^I(t) = \lambda^{\rm{eff}} (\adag + \a) (\ket{\Psi_+(t)}\bra{\Psi_+(t)} - \ket{\Psi_-(t)}\bra{\Psi_-(t)}) ,
\end{equation}
where the effective quantum coupling is~\footnote{In the JCM case the $B_{2k-1}$ term does not appear. However, this expression reduces to the correct form when the Floquet states are obtained from the semiclassical RWA Hamiltonian, because there is only one Fourier component in $B_{\pm}(t)$.} 
\begin{equation}
    \lambda^{\rm{eff}} = \lambda \sum_{k = -\infty}^{\infty} A_{2k}(B_{2k+1} + B_{2k-1}) .
\end{equation}
The effective equations of motion governing the field evolution are readily solved. Transforming back to the original frame, the solutions for the field states are given by
\begin{equation}
    \begin{split}
        \ket{\phi_{\pm}(t)} 
        &= e^{-i \omega_0 t \adag \a } e^{\pm 2 i \lambda^{\rm{eff}} \alphamag t}  \op{D}(\eta_{\pm}(t))  \ket{\phi_{\pm}(0)} ,
    \end{split}
    \label{field_state_fbrwa}
\end{equation}
where the time-dependent displacement of the field in phase space is $\eta_{\pm}(t) = \mp i e^{-i\phi} \lambda^{\rm{eff}} t$. This evolution of the coupled system may be interpreted as a dynamical polaron transformation, in which the field is displaced in a direction that depends on the state of the spin~\cite{Leggett1987, TwyIrish2005, TwyIrish2007}. Here the displacement amplitude is a function of time as well as coupling strength.

Together with the Floquet solutions for the spin, Eq.~\eqref{field_state_fbrwa} constitutes a closed-form analytical approximation for the time evolution of the joint spin--field state vector. In the following examples we focus on the excited-state probability $P(+z) = \abs{\langle +z \vert \Phi(t) \rangle}^2$ of the spin, a quantity with a clear connection to well-known results in quantum optics and significant relevance to quantum control. However, it is worth emphasizing that any dynamical quantity may be calculated from the state vector obtained within the FBRWA. 

Consider the initial state $\ket{\Phi(0)} = \ket{+z}\otimes\ket{\phi_0}$, with $\alpha = \bra{\phi_0} \a \ket{\phi_0}$ taken to be real. The general expression for the excited-state probability at later times is
\begin{equation}
\begin{split}
   P(+z) &= \abs{A_+(0)}^2 \abs{A_+(t)}^2 + \abs{B_-(0)}^2 \abs{B_-(t)}^2  \\
   & \quad + \big[A_+^*(0) B_-(0) A_+(t) B_-^*(t) e^{-i(q_+ - q_-)t} \\
   & \qquad \times \langle \phi_-(t) \vert \phi_+(t) \rangle + c.c. \big] .
\end{split}
\label{spin_prob}
\end{equation} 
The dynamics is determined by the Floquet coefficients and quasienergies, together with the inner product between the field states associated with different Floquet states. 

Let us first look at the Jaynes-Cummings model. The corresponding semiclassical model is exactly solvable. On resonance ($\Omega = \omega_0$), the quasienergies are $q_{\pm} = \pm(\tfrac{1}{2} \omega_0 + \lambda \alphamag)$ and the only non-zero coefficients of the Floquet modes are $A_0 = B_1 = 1/\sqrt{2}$. Hence $\lambda^{\rm eff} = \lambda/2$. Taking the initial state of the field to be a coherent state $\ket{\alpha}$, the excited-state probability is readily worked out to be  
\begin{equation}
       P(+z) = \tfrac{1}{2} + \tfrac{1}{2}e^{-\lambda^2 t^2/2} \cos(2 \lambda \alphamag t).
\end{equation}
This, of course, is the famous expression for sinusoidal Rabi oscillations with a Gaussian collapse envelope first obtained by Cummings \cite{Cummings1965}. In Cummings' approach, the time evolution of the joint state is given as a sum over the sinusoidal Rabi oscillations induced by each Fock state $\ket{n}$ contained in the initial coherent state. For large values of $\alphamag = \sqrt{\bar{n}}$, the weighting factor becomes sharply peaked around $\sqrt{\bar{n}}$ and the sum may be approximately evaluated, leading to the above expression. The collapse results from destructive interference of oscillations at incommensurate frequencies. 

Here we see a different physical interpretation of the collapse. The initial spin state is a superposition of the two semiclassical Floquet states. Within the FBRWA, the quantum interaction term does not cause transitions between the Floquet states; however, the evolution of the field depends on which Floquet state it is associated with. As the two field components become displaced in opposite directions, their inner product decreases, causing the collapse of the semiclassical Rabi oscillations. Gea-Banacloche showed similarly that the collapse occurs when the field states associated with two spin states become macroscopically distinguishable~\cite{GeaBanacloche1991, GeaBanacloche1992}. 

\begin{figure}[ht]
    \centering
    \includegraphics[width=1.0\linewidth]{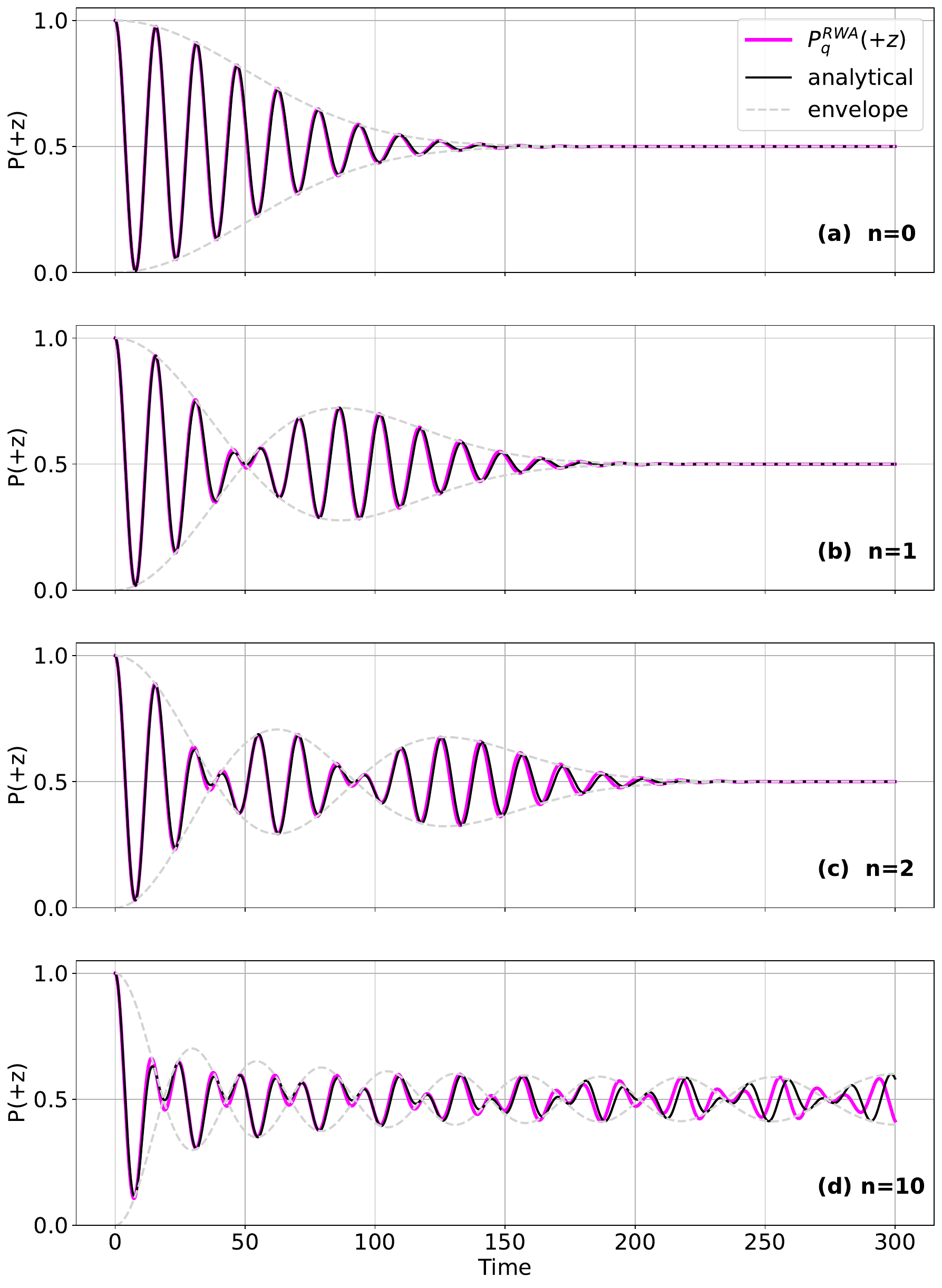}
    \caption{Collapse of Rabi oscillations for different initial field states in the JCM. The excited-state probability of the spin is plotted against time. Parameter values are $\alpha = 10$, $\lambda = 0.05$, and $\Omega = \omega_0 = 1$. Numerical solutions of the JCM Hamiltonian (magenta) are compared with Eq.~\eqref{fock_collapse_osc} (black); dashed grey lines indicate the collapse envelope from Eq.~\eqref{fock_collapse_osc}. The initial state of the system is $\ket{\Phi(0)} = \ket{+z} \otimes \ket{\alpha, n}$ with $n=$ (a) $0$, (b) $1$, (c) $2$, (d) $10$.}
    \label{fig:fock_collapse}
\end{figure}

Importantly, since Eq.~\eqref{field_state_fbrwa} determines the evolution of \emph{any} initial field states $\ket{\phi_{\pm}(0)}$, the analysis is not restricted to coherent states. For instance, consider an initial displaced Fock state $\ket{\alpha, n} = \op{D}(\alpha) \ket{n}$. Numerical studies of the dynamics have been carried out~\cite{Kim1990, Alexanian2022}, but no analytical results appear to have been previously reported. Using our approach, the only difference from the coherent-state case is in the field overlap and we easily obtain the expression 
\begin{equation}
        P(+z) = \tfrac{1}{2} + \tfrac{1}{2}e^{-\lambda^2 t^2/2} L_n(\lambda^2 t^2) \cos(2 \lambda \alphamag t).
\label{fock_collapse_osc}
\end{equation}
The Gaussian collapse envelope characteristic of a coherent field is modulated by the Laguerre polynomial $L_n(\lambda^2 t^2)$. Figure~\ref{fig:fock_collapse} illustrates this behavior. Rabi oscillations persist over longer times because the displaced Fock states $\ket{\alpha,n}$ are not minimum-uncertainty states, but have a width in phase space that increases with $n$. A larger relative displacement is required for the overlap between the two field states to become negligible; or, in other words, for the states to become macroscopically distinguishable. Nodes in the envelope result from destructive interference between the two displaced Fock states. 

Another interesting example is a superposition of two displaced Fock states:
\begin{equation}    
\ket{\phi_0} = (\ket{\beta, 0} + e^{-i \xi} \ket{\beta,1})/\sqrt{2}. \label{initial_sup}
\end{equation} 
For $\beta = \abs{\beta} e^{-i \phi}$ and $\xi = \phi$, the excited-state probability of the spin becomes
\begin{equation}
\begin{split}
    P(+z) = \tfrac{1}{2} + \tfrac{1}{2} e^{-\lambda^2 t^2/2} &\left[(1 - \tfrac{1}{2} \lambda^2 t^2) \cos(2 \lambda \abs{\beta} t) \right.\\
    &\left.- \lambda t \sin(2 \lambda \abs{\beta} t)\right]. \label{sup_collapse_osc}
\end{split}
\end{equation}
In this case the result cannot be simply expressed as a sinusoidal oscillation modified by a collapse envelope. Nevertheless, the analytical expression captures the dynamical behavior quite well (see Fig.~\ref{fig:sup_collapse} in the Appendix). 

Note that all the above results can be trivially extended to the off-resonance case. The only change required is to replace the Floquet quasienergies and states by the equivalent (also exact) solutions for $\Omega \neq \omega_0$. 

We now turn to the full Rabi model. The solutions have the same form when written in terms of the Floquet states, differing only in the value of $\lambda^{\rm eff}$. Finding the Floquet states of the Rabi model, however, is a challenging problem in itself. To illustrate the properties of the FBRWA solution with the Rabi interaction, we employ the approximate Floquet solutions for $\Omega = \omega_0$ given in \cite{Finney1994}, which are valid up to third order in $\epsilon \equiv \lambda \alphamag/(2 \Omega)$. 

\begin{figure}[t]
    \centering
    \includegraphics[width=1.0\linewidth]{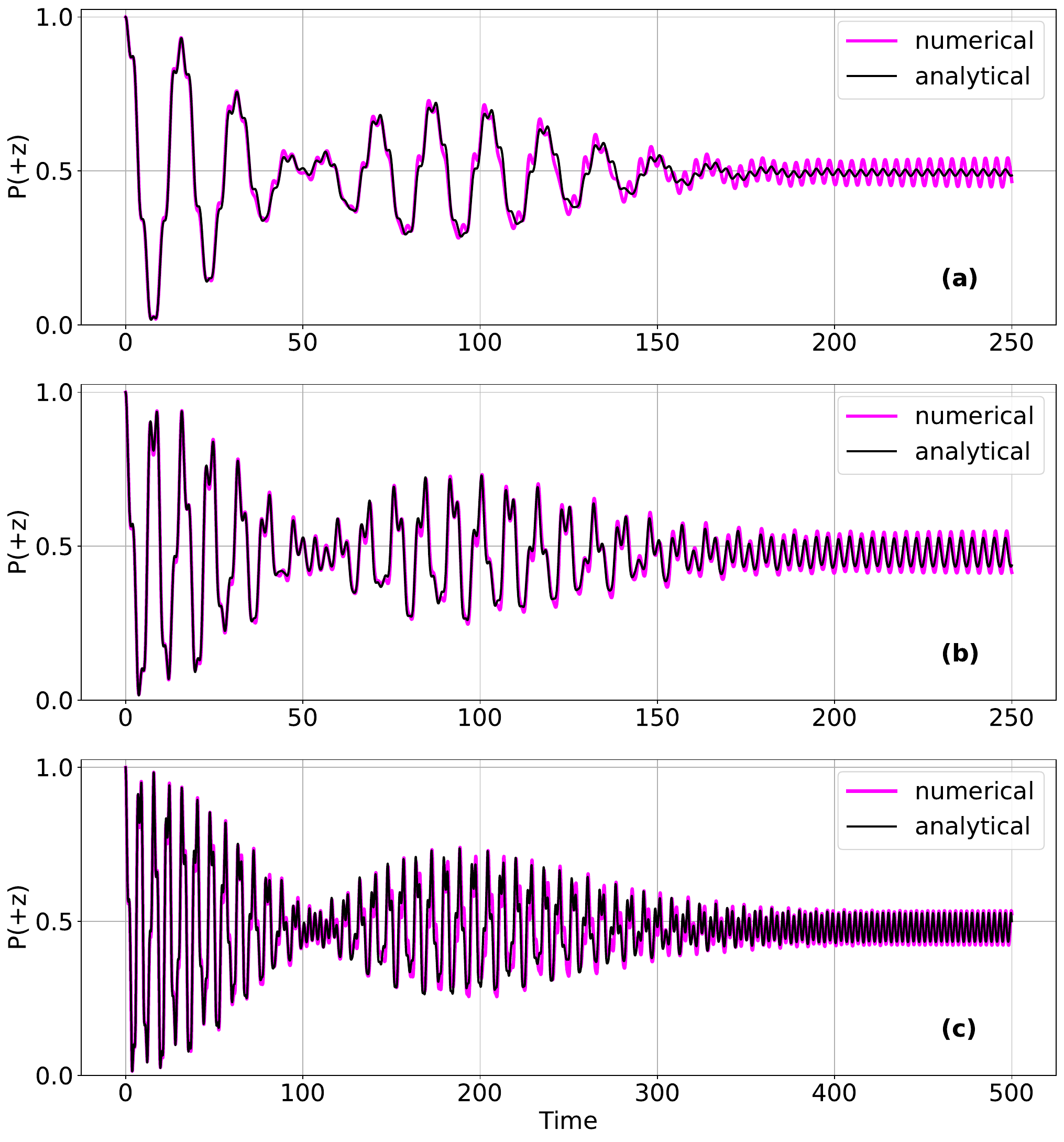}
    \caption{Collapse of Rabi oscillations in the resonant Rabi model ($\Omega = \omega_0 = 1$). The initial state is $\ket{+z}\otimes\ket{\alpha,1}$. Numerical solutions of the full Rabi Hamiltonian (magenta) are compared with the analytical solution (black) obtained from the FBRWA together with an analytical approximation for the Floquet states from \cite{Finney1994}. The parameters are (a) $\lambda = 0.02,\,\alpha = 10$, (b)  $\lambda = 0.02,\,\alpha = 20$ and (c) $\lambda = 0.01,\,\alpha = 40$.}
    \label{fig:Rabi_collapse}
\end{figure}

Figure~\ref{fig:Rabi_collapse} compares the population dynamics given by the analytical FBRWA expression in the Rabi model with a numerical solution of the full Hamiltonian. The field is initialized in the displaced Fock state $\ket{\alpha, 1}$. As the field components $\ket{\phi_{\pm}(t)}$ become displaced, the growing entanglement suppresses the coherence between the Floquet states that generates oscillations at the Rabi frequency. As in the JCM, the collapse is governed by the product of a Gaussian and a Laguerre polynomial. 

However, the collapse here is not complete; oscillations persist in the `quiescent' region, at a frequency near $2 \omega_0$~\cite{Zaheer1988, He2014}. The origin of this phenomenon is readily understood from Eq.~\eqref{spin_prob}. In the quiescent region, $\langle \phi_-(t) \vert \phi_+(t) \rangle \approx 0$. The spin dynamics is determined by $\abs{A_+(t)}^2$ and $\abs{B_-(t)}^2$, which represent the probability of $\ket{+z}$ in the Floquet states $\Psipket$ and $\Psimket$, respectively. In the JCM case, each of the Floquet coefficients has only one frequency component and hence the probabilities are constant. In the Rabi model, however, the counter-rotating terms cause the Floquet coefficients to develop additional components at higher frequencies. Since $A_+(t)$ ($B_-(t)$) contains only even (odd) multiples of $\omega_0$, the dominant correction involves cross terms with a frequency difference of $2\omega_0$. Consequently, the residual oscillations are unrelated to the quantum nature of the field. They arise because the Floquet states are time-dependent coherent superpositions of the bare spin eigenstates, a purely semiclassical effect.  Interestingly, a similar effect can be seen around the node of the $n=1$ Laguerre polynomial in the collapse envelope, for the same reason. 

The analytical curves plotted in Fig.~\ref{fig:Rabi_collapse} involve two approximations: the FBRWA itself, and the approximate Floquet solutions. The validity of each approximation depends on the parameters in different ways. In (a), the Floquet approximation is indistinguishable from the numerical solution of the semiclassical model, so differences between the two curves indicate limitations of the FBRWA. In the lower two panels, the larger value of $\epsilon$ creates inaccuracies in the Floquet approximation that have a visible effect at earlier times, especially in the central region of the bottom plot. By contrast, as $\alpha$ increases from the top panel to the bottom, the FBRWA improves, particularly for the residual oscillations in the collapse region.

As with any rotating-wave-type approximation, the validity of the FBRWA is predicated on a separation of timescales: the frequency of the rotating phase should be large compared to the associated transition matrix element. 
For the on-resonance JCM, this gives the criterion $q_+ - q_- - \omega_0 \gg \lambda/2$, which reduces to $\alphamag \gg 1/4$. Hence the approximation improves as the phase-space displacement $\langle \phi_0 \vert \a \vert \phi_0 \rangle$ of the initial field state is increased. The details differ for the off-resonant case and for the Rabi interaction (and in the latter case, time-dependent terms that are diagonal in the Floquet basis have also been neglected), but the essential scaling with $\alphamag$ holds, as seen from Fig.~\ref{fig:Rabi_collapse}. In general, the error in rotating-wave approximations also increases as time goes on~\cite{Angelo2005, Burgarth2022, Burgarth2024, Coleman2024}, which is true here as well.

One limitation of the FBRWA is that it predicts only the collapse of the Rabi oscillations, not their famous revivals. The interaction-induced displacement of the field components increases linearly with time, causing the overlap to asymptotically vanish for long times. Revivals, then, must originate from the terms in $H_q^I(t)$ that are off-diagonal in the Floquet states. This leads to the intriguing conclusion that the collapse and the revivals may be attributed to distinct physical processes in the Floquet basis. The collapse results from interaction terms that leave the semiclassical Floquet states unchanged but cause a state-dependent displacement of the field, generating entanglement between spin and field and thereby destroying the coherent oscillations between Floquet states. Revivals, on the other hand, are linked to interaction terms that drive transitions between the Floquet states. 

As we have established, the QCFD approach is quite general. We likewise expect that the FBRWA will prove useful for a range of models. Extension to an asymmetric~\cite{Armour2002,Braak2011,LarsonMavrogordatos} or anisotropic~\cite{Xie2014} Rabi model is simple: adding a bias term to the spin will alter the Floquet solutions, while allowing different coupling constants for the co-rotating and counter-rotating terms will change $\lambda^{\rm{eff}}$. Multilevel systems and spin chains can also be treated within this formalism, opening up a systematic approach to studying Floquet engineering with quantum fields~\cite{Bomantara2016,Sentef2020,Li2020,Eckhardt2022,Perez-Gonzalez2024}. Multiphoton interactions will be more complicated to address but intriguing to explore from this perspective. 

The framework presented here opens up broad prospects for future work, both within and beyond the FBRWA. Closed-form analytic expressions are readily derived from the FBRWA yet accurately capture even quite complicated short-time dynamics. Such computational and conceptual simplicity lends itself to assessing quantum field effects on control protocols deployed in quantum technologies. As we have demonstrated using the Jaynes-Cummings and Rabi models, this approach provides physical insight that draws various aspects of dynamics into a cohesive picture. Both the practical applications and the foundational implications for how classical and semiclassical behavior emerge from quantized fields offer tantalizing avenues for exploration.

EKT thanks J. Gea-Banacloche for helpful conversations in the early stages of this work, and H.F.A. Coleman, M. Spittle, and R. Morrison for lively discussions. ADA acknowledges support from a Leverhulme Trust Research Project Grant (RPG-2023-177). 

The data that support the findings of this article are openly available\,\cite{Open_Data_Twyeffort_2025}.

\bibliography{main}

\newpage
\appendix
\section{Appendix}
\section{Derivation of specific expressions for the JCM and quantum Rabi model}
Here we demonstrate how the Hamiltonian of Eq.~\eqref{H_orig} is separated into semiclassical and quantum terms and derive the specific forms of $H_{sc}(t)$ and $H_q^I(t)$. Following the `recipe' of \cite{TwyIrish2022} for obtaining the semiclassical limit, two transformations are carried out. First, the Hamiltonian is written in the interaction picture with respect to the field, using the operator $\op{U}(t) = e^{-i \omega_0 t \adag \a}$:
\begin{align}
    H^I(t) = \tfrac{1}{2} \Omega\sigmaz &+ \lambda \left[f(e^{-i \omega_0 t} \a , e^{i \omega_0 t} \adag) \sigmap \right. \nonumber\\
    &+\left.f^{\dag}(e^{-i \omega_0 t} \a, e^{i \omega_0 t} \adag) \sigmam\right].
\end{align}
Next, a displacement transformation $\op{D}(\alpha) = \exp(\alpha\adag - \alpha^* \a)$ with $\alpha \equiv \alphamag e^{-i\phi}$ is applied, resulting in a transformed Hamiltonian that can be expressed as a sum of two terms:
\begin{equation}
        \tilde{H}^{I}(t) = H_{sc}(t)\otimes \op{I}_f + H_q^I(t) ,
\label{displ_transform}
\end{equation}
where 
\begin{align}
H_{sc}(t)= \tfrac{1}{2} \Omega \sigmaz &+ \lambda\left[f(\alphamag e^{-i (\omega_0 t+\phi)} , \alphamag e^{i (\omega_0 t+\phi)} ) \sigmap\right. \nonumber \\
&+ \left.f^*(\alphamag e^{-i (\omega_0 t+\phi)} ,  \alphamag e^{i (\omega_0 t+\phi)}) \sigmam\right]   
\end{align}
is the semiclassical equivalent of Eq.~\eqref{H_orig} and 
\begin{equation}
H_q^I(t) \equiv \lambda [f(e^{-i \omega_0 t} \a, e^{i \omega_0 t} \adag) \sigmap + f^{\dag}(e^{-i \omega_0 t} \a, e^{i \omega_0 t} \adag) \sigmam]
\label{HqI}
\end{equation}
is the quantum interaction term. 

Now the quantum interaction $H_q^I(t)$ can be expressed in the Floquet basis $|\Psi_{\pm}(t)\rangle$. For the JCM,  $f(e^{-i \omega_0 t} \a, e^{i \omega_0 t}\adag) = e^{-i \omega_0 t}\a$, and the first term of Eq.~\eqref{HqI} takes the form 
\begin{widetext}
\begin{equation}
    \begin{split}
        \lambda e^{-i \omega_0 t} \a \sigmap = \lambda \a \sum_{k,l = -\infty}^{\infty} \big[&A_{2k} B_{2l+1} e^{2(l-k)i\omega_0 t} (\Psipket \Psipbra - \Psimket \Psimbra) \\
        &  + e^{-i(q_+ - q_-)t} B_{2k+1} B_{2l+1} e^{(2k+2l+1)i \omega_0 t} \Psimket \Psipbra \\
        & - e^{i(q_+ - q_-)t} A_{2k} A_{2l} e^{-(2k+2l+1)i \omega_0 t} \Psipket \Psimbra  \big] .
    \end{split} \label{a_sigmap}
\end{equation}
The quantum Rabi model additionally contains the counter-rotating term
\begin{equation}
   \begin{split}
       \lambda e^{i \omega_0 t} \adag \sigmap = \lambda \adag \sum_{k,l = -\infty}^{\infty} \big[&A_{2k} B_{2l+1} e^{2(l-k+1)i\omega_0 t} (\Psipket \Psipbra - \Psimket \Psimbra) \\
       &  + e^{-i(q_+ - q_-)t} B_{2k+1} B_{2l+1} e^{(2k+2l+3)i \omega_0 t} \Psimket \Psipbra \\
       & - e^{i(q_+ - q_-)t} A_{2k} A_{2l} e^{-(2k+2l+3)i \omega_0 t} \Psipket \Psimbra  \big] .
   \end{split} \label{adag_sigmap}
\end{equation}
\end{widetext}
In this basis, $H_q^I(t)$ appears as a complicated function of time with both diagonal and off-diagonal terms in the Floquet states, making the coupled differential equations~\eqref{field_diffeqs} for the field difficult to solve.

Following the same reasoning that justifies dropping the counter-rotating terms from the Rabi model to obtain the simpler JCM, we argue that the dominant contributions in Eqs.~\eqref{a_sigmap} and \eqref{adag_sigmap} arise from the time-independent terms. The first line of each equation, which is diagonal in the Floquet states, possesses zero-frequency components for all $l=k$. The off-diagonal terms, however, exhibit an additional time dependence arising from the quasienergy difference $q_+ - q_-$. Zero-frequency components are only found at isolated resonances satisfying the condition $q_+ - q_- = (2k + 2l + 1) \omega_0$. Away from any such resonances, the interaction terms that drive transitions between the Floquet states are suppressed relative to the time-independent diagonal terms and may be neglected to lowest order. This is the FBRWA analyzed in the main text. 

Within the range of parameters considered in the examples shown, we have verified that the lowest frequency at which a resonance, and hence a zero-frequency contribution from the off-diagonal Floquet terms, occurs well outside the validity of the approximate Floquet solutions we have employed. However, it should be noted that such resonances may need to be taken into account for other parameter regimes. Exploring their effect on the dynamics of the coupled system and the validity of the FBRWA is an open question for future work.

\section{Rabi oscillation collapse for superposition states}

\begin{figure}[ht]
    \centering
    \includegraphics[width=1.0\linewidth]{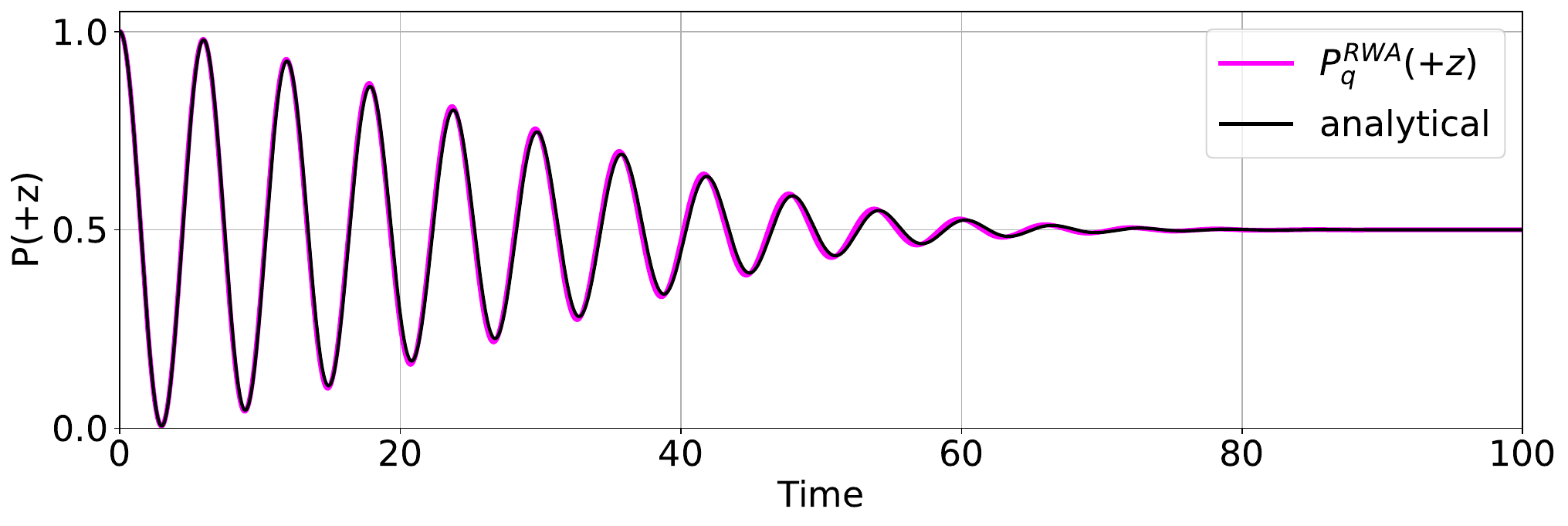}
    \caption{Collapse of Rabi oscillations with the field initialized in the superposition state given in Eq.~\eqref{initial_sup}. The excited-state probability of the spin is plotted as a function of time. Parameter values are $\beta = 10$, $\lambda = 0.05$, and $\Omega = \omega_0 = 1$. Numerical solution of the JCM Hamiltonian (magenta) is compared with the analytical expression given in Eq.~\eqref{sup_collapse_osc} (black).}
    \label{fig:sup_collapse}
\end{figure}

Figure ~\ref{fig:sup_collapse} illustrates the dynamics induced by a field initialized in the superposition of displaced Fock states given in Eq.~\eqref{initial_sup}. The FBRWA expression for the excited-state spin population [Eq.~\eqref{sup_collapse_osc}] agrees extremely well with a full numerical integration of the JCM, capturing all of the Rabi oscillation collapse. 

\end{document}